\documentclass[reprint, amsmath, amssymb, aps, pre, twocolumn]{revtex4-2}

\usepackage{graphicx} 
\usepackage{dcolumn}  
\usepackage{bm}       
\usepackage{hyperref} 
\usepackage{xcolor}   
\usepackage{url}
\usepackage{booktabs}
\usepackage[table]{xcolor}
\usepackage[utf8]{inputenc}
\usepackage{tabularx}
\usepackage{booktabs}

\begin{document}

\title{A Network Inefficiency Metric for Structural Stress Detection in Hedera Transactions}

\author{Deep Nath}
\email{d.nath@exp.science}

\author{Paolo Tasca}
\email{p.tasca@exp.science}

\author{Nikhil Vadgama}
\email{nikhil.vadgama@exp.science}

\author{Marco Alberto Javarone}
\email{m.javarone@exp.science}

 \affiliation{Exponential Science Foundation, Lugano, Switzerland}

\begin{abstract}
Quantifying structural stress in transaction networks requires metrics that capture structural organization beyond transaction volume alone. In this work, we introduce the Inefficiency Metric, a deterministic indicator designed to characterize the routing structure of capital flows in decentralized systems. Using Principal Component Analysis and Pearson correlation matrices computed from a six-year Hedera transaction dataset, we identify two dominant and largely independent structural dimensions: the effective diameter, related to the spatial extension of transaction propagation, and the closeness centrality, associated with the efficiency of network-level flow processing.
The proposed metric reveals significant topological fluctuations associated with major macroeconomic and ecosystem-level events. Increased inefficiency is observed during periods marked by intermediary fragmentation or rapid smart-contract expansion, whereas lower inefficiency corresponds to phases of network compaction during market stress or institutional concentration. Comparison with a seven-dimensional Isolation Forest approach shows that the metric effectively captures severe multidimensional anomalies while preserving a clear structural interpretation. Overall, these results provide a physics-inspired framework for relating the large-scale organization of decentralized transaction networks to observable economic dynamics.
\end{abstract}

\maketitle

\section{Introduction}

Distributed Ledger Technologies (DLTs) are changing the way money and information move around the world digitally \cite{Nakamoto2008bitcoin}. Different DLTs adopt different consensus algorithms to achieve consistency, fault tolerance, and transaction validation in decentralized environments \cite{Lashkari2021BlockchainConsensus}. Early DLTs such as Bitcoin use Proof of Work (PoW)-based consensus \cite{Sapra2023PoWEnvironment} over a linear chain structure. However, newer architectures employ mechanisms such as Proof of Stake (PoS) \cite{Saleh2021ProofOfStake}, Directed Acyclic Graphs (DAGs) \cite{Raikwar2024DAGConsensus}, and asynchronous Byzantine Fault Tolerant (aBFT) \cite{Tholoniat2022BlockchainBFT} protocols to improve scalability and parallel transaction processing \cite{Akhtar2019BlockchainHashgraph}.

Among several new-generation DLTs~\cite{javarone03}, this study focuses on Hedera \cite{Baird2016hashgraph}. Unlike traditional linear blockchains, whose topology is constrained by sequential block production, Hedera combines a DAG with aBFT \cite{Baird2020}. This architecture generates a complex, non-linear transaction network where accounts become vertices and value flows become edges \cite{Amherd2023centralised}. Yet, despite this structural complexity~\cite{javarone02} and high enterprise throughput, the topological evolution of the Hedera network remains largely unexplored. More in detail, how the underlying topology of the transaction networks behaves and evolves under severe macroeconomic stress has not been formally established.
Network science~\cite{barabasi2013,Newman2003ComplexNetworks,caldarelli2019} provides a robust framework for mapping these interactions through topological analysis \cite{Amherd2023centralised, Saengchote2023DeFiTopology}.

However, analyzing DLTs presents a unique challenge: they are highly dynamic, multidimensional systems that generate vast amounts of data. Their evolution can be investigated through traditional financial measures such as transaction volume, Total Value Locked (TVL), or market capitalization \cite{Metelski2022DeFi}. Yet, these metrics fail to capture the network structure, e.g. obscuring how users and institutions actually interact on-chain. 
Financial indicators frequently decouple from true network utility, remaining susceptible to speculative noise \cite{Inuduka2024NoiseTrader}, wash trading \cite{Tosic2025WashTrading}, and concentrated capital movements that inflate volume without broadening participation \cite{Barnes2018Cryptocurrency}. Thus, to assess the health of a decentralized ecosystem, we must move beyond surface-level financial data and directly measure its physical topology.

Also, although machine learning models are usually very useful for processing high-dimensional data \cite{Bishop2006PRML}, when applied to DLT transaction networks are often unable to detect the difference among temporary micro-noise—like bot activity, wallet reorganizations, or routine liquidity sweeps—and significant, network-wide structural changes \cite{Alsagheer2023MLGovernance}. 

Therefore, to track the macroeconomic health of a DLT network, we suggest a different approach able to reduce this complex data into a single and interpretable metric.

To this end, in this work, we focus on the structural tension inherent in the Hedera network: the trade-off between spatial expansion and internal accessibility. When this network faces macroeconomic shocks or periods of intense smart-contract activity, transaction pathways stretch---capital must travel longer distances. 
Conversely, during market panics or institutional consolidation, activity compacts, routing directly through a few centralized hubs. Capturing this push-and-pull requires isolating the right topological dimensions. To quantify this tension, we introduce a new indicator: the Inefficiency Metric ($\mathcal{I}$). Using Principal Component Analysis (PCA) and correlation matrices, we reduce the topological feature space of the Hedera network to identify independent structural dimensions. Notably, we restrict the metric to effective diameter ($\mathcal{D}_{eff}$) and closeness centrality ($\langle \mathcal{C} \rangle$), measuring the friction between spatial stretch and core accessibility.

To validate the performance of the proposed metric in detecting genuine macro-events, we compare its output against an unsupervised Isolation Forest model operating across seven topological dimensions. 
Remarkably, we find that the Inefficiency Metric allows us to successfully identify known macroeconomic events in Hedera transaction history—such as market crashes, exchange collapses, and institutional entry points—directly from the topology.

The remainder of this paper is organized as follows: Section~\ref{sec:methods} details the network construction and the Inefficiency Metric. Section~\ref{sec:results} presents results from applying the metric to Hedera. Finally, in Section~\ref{sec:conclusion}, we summarise the main findings and suggest future developments.

\section{Methodology}\label{sec:methods}
To capture the structural evolution of the Hedera network, we model the weekly transaction data as directed networks. To this end, let $\mathcal{G}(\mathcal{V},\mathcal{E})$ represent the Hedera weekly transaction networks (HWTN). Hereinafter, we consider the largest weakly connected component (LWCC) for a comprehensive understanding of a transaction network. $\mathcal{V}$ and $\mathcal{E}$ denote the set of vertices (user accounts) and edges (directed transactions), respectively, in the LWCC. Filtering out smaller, isolated fragments ensures that subsequent metrics reflect a cohesive economic structure rather than a collection of disconnected components. 
We use the resulting directed graphs for measuring explicit capital flow, path length, and routing efficiency, as financial transactions are strictly unidirectional. 
However, in some cases, we also use the undirected version of HWTNs to assess the structural inequality of hierarchy. For these metrics, we focus on the overall connectivity and structural relationships between accounts, regardless of the specific direction of transaction flow.

To evaluate the structural evolution of the Hedera transaction network, we use some topological metrics that capture distinct physical properties of capital routing. 
While traditional financial indicators, such as transaction volume or token price, might not fully describe how users and accounts are structurally connected within the network, network science offers more metrics to quantify macroscopic spatial expansion, internal routing efficiency, and structural hierarchy. 
These metrics translate the complex interactions of millions of accounts into interpretable macro-states. By doing so, they provide a mathematical lens to identify distinct structural shifts, such as extreme decentralization or network compaction, and correlate them directly with real-world macroeconomic events. 
In graph theory, the diameter ($\mathcal{D}$) is calculated as the longest shortest-path distance between any two reachable nodes \cite{Borner2007NetworkScience}. While this metric theoretically represents the longest possible transaction route, it is highly sensitive to topological outliers. In large-scale ledgers, a single anomalous chain of trivial transactions can artificially inflate $\mathcal{D}$, causing the metric to severely misrepresent the actual macroscopic structure of the network. Therefore, to quantify the macroscopic spatial stretch of the network more effectively, we compute the $90^{th}$ percentile effective diameter ($\mathcal{D}_{eff}$). $\mathcal{D}_{eff}$ measures the minimum number of directed hops required for capital to route between 90\% of all reachable accounts. 

Because exact all-pairs shortest-path computation scales superlinearly and is intractable for millions of nodes, we employ a Monte Carlo sampling approach. We generate $M$ independent random ensembles, each containing $N = 1000$ source nodes, $\mathcal{V}_{sample,m} \subseteq \mathcal{V}$. 
For each ensemble $m$, we use a Breadth-First Search to calculate the minimum directed distance, $d(i,j)$, to all reachable nodes $j$. We aggregate valid path lengths into an ensemble distribution $P_m$

\begin{equation}
P_m = \{ d(i, j) \mid i \in \mathcal{V}_{sample,m},\ j \in \mathcal{V},\ d(i,j) > 0 \}.
\end{equation}

Self-distances are excluded to prevent artificial downward bias. The effective diameter for ensemble $k$ is defined using the empirical cumulative distribution function $F_{P_m}(d)$

\begin{equation}
\mathcal{D}_{eff,m} = \inf \{ d \in \mathbb{R} \mid F_{P_m}(d) \geq 0.90 \}.
\end{equation}

Then, we obtain the $90^{th}$ percentile of the effective diameter as

\begin{equation}
{\mathcal{D}}_{eff} = \frac{1}{M} \sum_{m=1}^{M} \mathcal{D}_{eff,m} .
\end{equation}
Theoretically, $\mathcal{D}_{eff}$ represents the macroscopic spatial stretch of the ledger, quantifying the maximum number of directed transactional hops required to successfully route capital across $90\%$ of the active economic network. However, to assess how efficiently liquidity moves through the ecosystem, we also evaluate various other metrics.
Among these measures, closeness centrality ($\mathcal{C}_i$) quantifies how efficiently the $i^{th}$ node can reach all other reachable nodes in the graph \cite{Newman2003ComplexNetworks}. In a transaction network, $\mathcal{C}_i$ provides information about how rapidly a specific node can broadcast or transfer capital across the broader ecosystem. For a node $i$ and its set of reachable nodes $R_i$, we define it as

\begin{equation}
\mathcal{C}_i = \frac{|R_i|-1}{\sum_{j \in R_i \setminus \{i\}} d(i,j)}.
\end{equation}

Nodes with high values occupy structurally dominant routing positions. However, because our study focuses on macroscopic phase states rather than individual nodes, we compute the average closeness centrality across the network

\begin{equation}
\langle \mathcal{C} \rangle = \frac{1}{|\mathcal{V}|} \sum_{i \in \mathcal{V}} \mathcal{C}_i.
\end{equation}

In graph theory, this average value signifies the overall topological compactness and global routing efficiency of the system. In the specific context of the Hedera transaction network, $\langle \mathcal{C} \rangle$ directly measures how easily capital flows through the system. It highlights periods when transactions can move efficiently across the ledger without being delayed by long or disconnected pathways.

We evaluate the average clustering coefficient, $C$, to quantify the formation of closed triadic trading loops, which represent localized arbitrage cycles or decentralized exchange routing \cite{Ducruet2014Spatial}. As the transaction network is directed, local clustering coefficient, $C_i$, accounts for both the total degree and reciprocal connections to evaluate the true fraction of realized triangles. We obtain

\begin{equation}
C_i = \frac{T_i}{k^{tot}_i(k^{tot}_i - 1) - 2k^{\leftrightarrow}_i}, \quad C_{av} = \frac{1}{|\mathcal{V}|} \sum_{i \in \mathcal{V}} C_i.
\end{equation}

\noindent here $T_i$ is the number of directed triangles involving node $i$, $k^{tot}_i$ is the total degree (the sum of in-degree and out-degree), and $k^{\leftrightarrow}_i$ represents the bilateral (reciprocal) degree of node $i$.
We also calculate the Gini coefficient, $G$, associated with undirected version of $\mathcal{G}(\mathcal{V},\mathcal{E})$ \cite{Badham2013Degree}. This coefficient measures structural inequality within the undirected degree distribution. So, given $k_1 \leq k_2 \leq \dots \leq k_n$ the sorted degrees of the network,  
$G$ is calculated as

\begin{equation}
G = \frac{2\sum_{i=1}^{n} i k_i}{n\sum_{i=1}^{n} k_i} - \frac{n+1}{n}
\end{equation}

\noindent where $n=|\mathcal{V}|$. $G = 0$ indicates perfect equality (all accounts possess the same number of connections) and $G \to 1$ indicates maximal concentration (a single node monopolizes all connections). In the context of the Hedera transaction network, $G$ serves as a direct measure of economic centralization and hierarchical dominance. High $G$ reveals a heavily centralized, ``hub-and-spoke" topology where a small subset of accounts---such as centralized exchange hot wallets, massive enterprise smart contracts, or native staking hubs---processes the vast majority of all transactions. Conversely, the low value of $G$ reflects a flattened, highly decentralized landscape characterized by widespread peer-to-peer retail activity or evenly distributed application usage. By tracking $G$ over time, we can identify whether network activity is spreading evenly across many regular users or concentrating around a few massive institutional hubs.

On the other hand, degree assortativity ($r$) quantifies the topological mixing patterns of the network by measuring the tendency of nodes to connect with others of similar degree~\cite{Newman2003ComplexNetworks,javarone01}. For an undirected network with $|\mathcal{E}|$ edges connecting nodes of degrees $k_i$ and $k_j$, the assortativity coefficient is defined as

\begin{equation}
r= \frac{[\sum_{\mathcal{E}_{ij} \in \mathcal{E}} k_i k_j] - [\sum_{i \in \mathcal{V}} \frac{k_i^2}{2}]^2 / |\mathcal{E}|}{[\sum_{i \in \mathcal{V}} \frac{k_i^3}{2}] - [\sum_{i \in \mathcal{V}} \frac{k_i^2}{2}]^2 / |\mathcal{E}|}
\end{equation}

In the context of the Hedera transaction network, $r$ serves as a critical indicator of economic routing behavior, distinguishing between institutional trading loops and retail-driven service topologies, and ranging in $-1 \le r \le 1$. $r > 0$ indicates that highly connected accounts---such as centralized exchanges, enterprise treasury accounts, or major liquidity pools---preferentially transact with one another. This structural state reflects closed-loop institutional clearing or inter-corporate ecosystem activity. Conversely, $r < 0$ entails a disassortative, star-like topology. This occurs when a few massive central hubs (e.g., a centralized exchange hot wallet or a popular smart contract) act as primary service providers for thousands of isolated, low-degree retail accounts. Thus, through $r$, we can determine whether the dominant network volume is circulating exclusively among elite institutional players or bridging outward to the broader retail public.

We also compute the the average degree, $\langle k \rangle$, and density, $\rho$, associated with directed $\mathcal{G}(\mathcal{V},\mathcal{E})$. These measure the overall transactional volume per user and the utilization rate of possible market connections, respectively

\begin{equation}
\langle k \rangle = \frac{|\mathcal{E}|}{|\mathcal{V}|}, \quad \rho = \frac{|\mathcal{E}|}{|\mathcal{V}|(|\mathcal{V}|-1)}.
\end{equation}

Collectively, these measures can be considered as topological features to establish a profile of the Hedera network. 
While effective diameter and closeness centrality capture the physical routing stretch and flow efficiency, metrics such as the Gini coefficient, and assortativity independently quantify the ecosystem structural hierarchy and institutional consolidation. By evaluating these distinct topological dimensions, we construct the mathematical foundation for a novel, network-based inefficiency metric, allowing us to accurately detect and isolate periods of acute structural stress from raw transactional data.

Evidently, multidimensional, interacting topological features govern the structural evolution of the Hedera transaction network. Therefore, we employ Principal Component Analysis (PCA) over the encompassing topological feature space~\cite{Jolliffe2002PCA}. PCA is fundamentally variance-driven and highly sensitive to the relative scale of input features. In the context of the HTWN, the topological metrics may inhabit vastly different scales. Prior to eigen-decomposition, the time-series data for all structural metrics was standardized (Z-score standardization). Each time-series vector was independently transformed to possess a mean of zero and a unit variance

\begin{equation}
z_{i} = \frac{x_{i} - \mu}{\sigma}
\end{equation}

\noindent where $x_i$ is the raw topological observation, and $\mu$ and $\sigma$ are the respective mean and standard deviation of that specific feature over the 6-year timeline. 
This transformation projects all variables into a scale-invariant phase space, ensuring that the PCA identifies principal components based entirely on structural covariance rather than arbitrary mathematical magnitude.
We project the standardized high-dimensional space onto its first two principal components (PC1 and PC2). This captures the dominant variance of the system temporal evolution. To interpret the physical meaning of these dimensions, we map the original structural metrics as dimensional vectors within the reduced phase space. The geometric arrangement of these loading vectors provides a rigorous foundation for variable selection: vectors that align at $180^\circ$ angles represent inversely coupled systemic behaviors, whereas vectors operating orthogonally (at $90^\circ$ angles) represent fundamentally independent, uncoupled topological features.
As complex systems natively generate correlated macro-states, the selected topological features might capture overlapping structural information. To prevent algorithmic bias, we first standardize the data with Z-score standardization, as mentioned above. We then compute the Pearson correlation matrix to quantify linear covariance between any two standardized network variables $x$ and $y$ \cite{Pearson1895Correlation}

\begin{equation}
\sigma_{xy} = \frac{\sum_{i=1}^{n} (x_i - \bar{x})(y_i - \bar{y})}{\sqrt{\sum_{i=1}^{n} (x_i - \bar{x})^2} \sqrt{\sum_{i=1}^{n} (y_i - \bar{y})^2}}
\end{equation}

\noindent where $\bar{x}$ and $\bar{y}$ represent the respective means of the time series. Also, $-1 \le \sigma \le +1$. Crucially, for the purpose of dimensionality reduction, we consider only the absolute magnitude of $|\sigma|$. Two topological features can be considered to evolve in synchronized and linear way if $|\sigma| \to 1$---either directly ($\sigma \to +1$) or inversely ($\sigma \to -1$). 

Complex systems, such as DLTs, natively generate highly correlated macro-states. If two metrics exhibit $|\sigma| \approx 1.0$, they describe the exact same physical structural constraint. If we do not exclude linearly dependent metrics, they can skew our analysis associated with isolation forest. Therefore, the Pearson matrix serves as a necessary diagnostic filter, allowing us to objectively identify collinear variables.
We introduce a dimensionless Topological Inefficiency Metric ($\mathcal{I}$) to quantify macroscopic routing degradation. Since $\mathcal{D}_{eff}$ and $\mathcal{C}$ inhabit vastly different numerical scales, we apply Min-Max normalization across the timeline, yielding scaled variables $\tilde{\mathcal{D}}_{eff}(t)$ and $\tilde{\mathcal{C}}(t)$ bounded strictly between $0$ and $1$. Then, the Topological Inefficiency Metric score is defined as

\begin{equation}
\mathcal{I}(t) = \tilde{\mathcal{D}}_{eff}(t) - \tilde{\mathcal{C}}(t).
\end{equation}
\noindent Notice that this metric bounds between $-1$ and $+1$. A score approaching $+1$ indicates severe topological congestion where the network is maximally stretched and its core is highly inaccessible. Conversely, a score approaching $-1$ denotes an optimally compact, easily traversable state.
To rigorously validate whether $\mathcal{I}(t)$ accurately captures true structural stress within the Hedera ledger, we employ an independent and unsupervised machine learning approach, utilizing the broader multidimensional feature space. If $\mathcal{I}(t)$ is a robust metric, its extreme spikes should natively align with the high-dimensional anomalies detected by the algorithm.
To this end, we use an Isolation Forest, an ensemble-based anomaly detection algorithm uniquely suited for identifying structural outliers without relying on predefined density assumptions \cite{Liu2012IsolationBased}. Unlike standard algorithms that attempt to model a `normal' baseline, the Isolation Forest explicitly isolates anomalous observations by randomly selecting features and assigning random split values. Because anomalous network states, such as severe topological collapses, centralized exchange failures, or massive transaction injections, deviate sharply from typical evolutionary baselines, they require significantly fewer random partitions to be isolated. 
All temporal data is standardized to zero mean and unit variance to prevent variables with larger absolute magnitudes from skewing the partition distance. To prevent bias in establishing an anomaly rate, the model adopts an autonomous contamination threshold, allowing the algorithm to determine the natural boundary of structural outliers. Consequently, weekly network states that produce notably short average path lengths across the ensemble of decision trees are flagged as critical structural anomalies (labeled as $-1$ within the phase space). By superimposing these ML-derived, high-dimensional anomalies onto the timeline of our deterministic $1$-dimensional Inefficiency score, we independently verify the diagnostic precision and reliability of the proposed metric.
\begin{figure}
    \centering
     \includegraphics[width=\columnwidth]{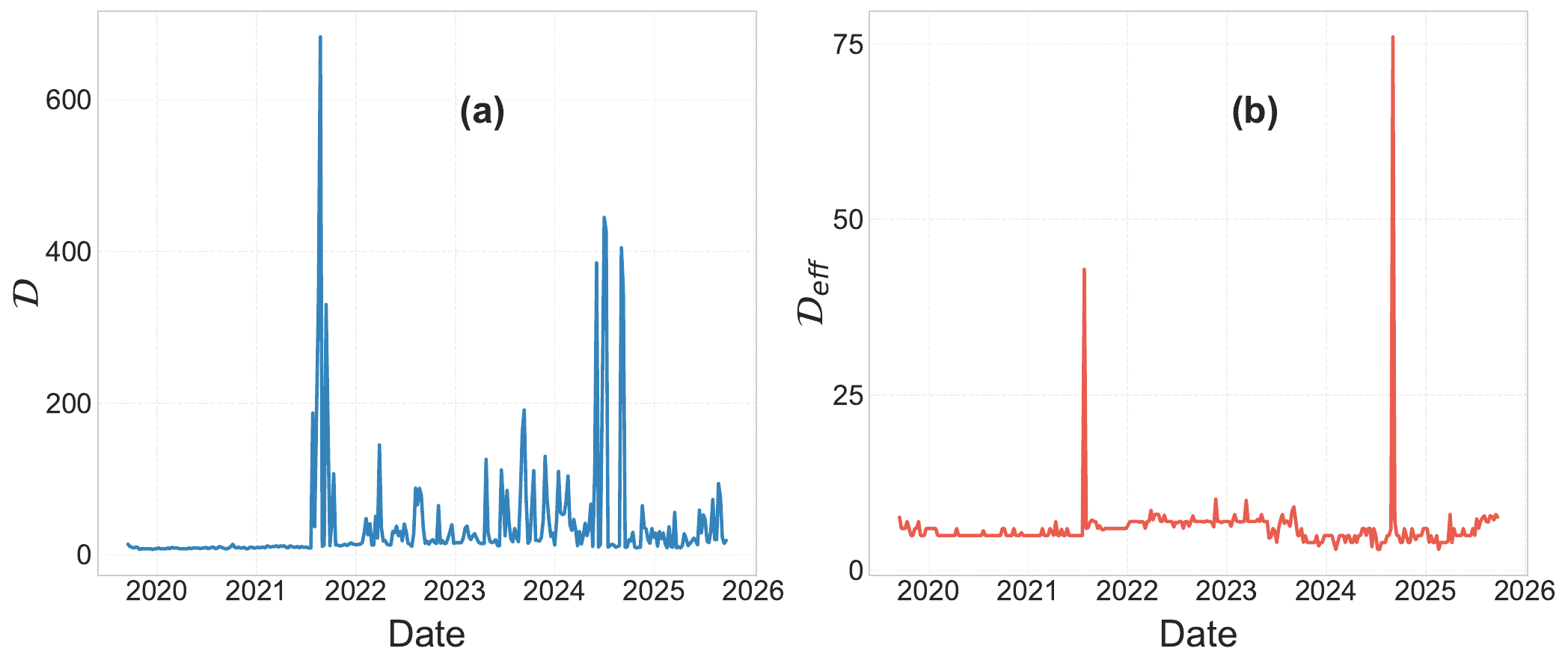}
    \caption{Temporal evolution of the (a) diameter, $\mathcal{D}$, and, (b) $90^{th}$ percentile effective diameter, $\mathcal{D}_{eff}$, within the Hedera transaction network. The extreme spikes--reaching over $600$--show that $\mathcal{D}$ is too easily skewed by isolated, unusual transaction chains. This perfectly justifies our decision to use $\mathcal{D}_{eff}$ instead.}
    \label{fig:diameter_dual}
\end{figure}

\section{Results}\label{sec:results}

Figure~\ref{fig:diameter_dual}(a) shows the time evolution of the network diameter, $\mathcal{D}$, for HWTNs. As the Hedera transaction network evolves, $\mathcal{D}$ periodically experiences massive and artificial inflations, frequently exceeding $600$ directed topological hops within a single week. In the economic reality of a distributed ledger, these transient spikes do not signify genuine systemic routing friction or critical structural stress. 
Instead, they expose the vulnerability of absolute longest-path metrics to extreme topological outliers, such as an isolated smart contract execution or a bot-driven account generating a long, linear chain of trivial transfers. 
If this raw, unpruned diameter were incorporated into our unsupervised Isolation Forest model or our deterministic $1$D metric, this localized noise would trigger severe false-positive anomalies, fundamentally obscuring the true macroscopic state of the network. By visualizing how a single anomalous chain can artificially inflate the perceived spatial stretch of the ledger, this distribution proves why standard $\mathcal{D}$ must be discarded in favor of $\mathcal{D}_{eff}$. In Figure~\ref{fig:diameter_dual}(b), we show the time evolution of the $90^{th}$ percentile effective diameter, $\mathcal{D}_{eff}$. Here, we notice that the volatility associated with $\mathcal{D}_{eff}$ is far less compared to $\mathcal{D}$. Therefore, we use $\mathcal{D}_{eff}$ within the foundation of our analyses. This substitution guarantees that our measurement of structural stress remains anchored to the actual liquidity pathways utilized by the vast majority of the ecosystem, rather than being influenced by peripheral network artifacts.

\begin{figure}[h]
    \centering
    \includegraphics[width=\columnwidth]{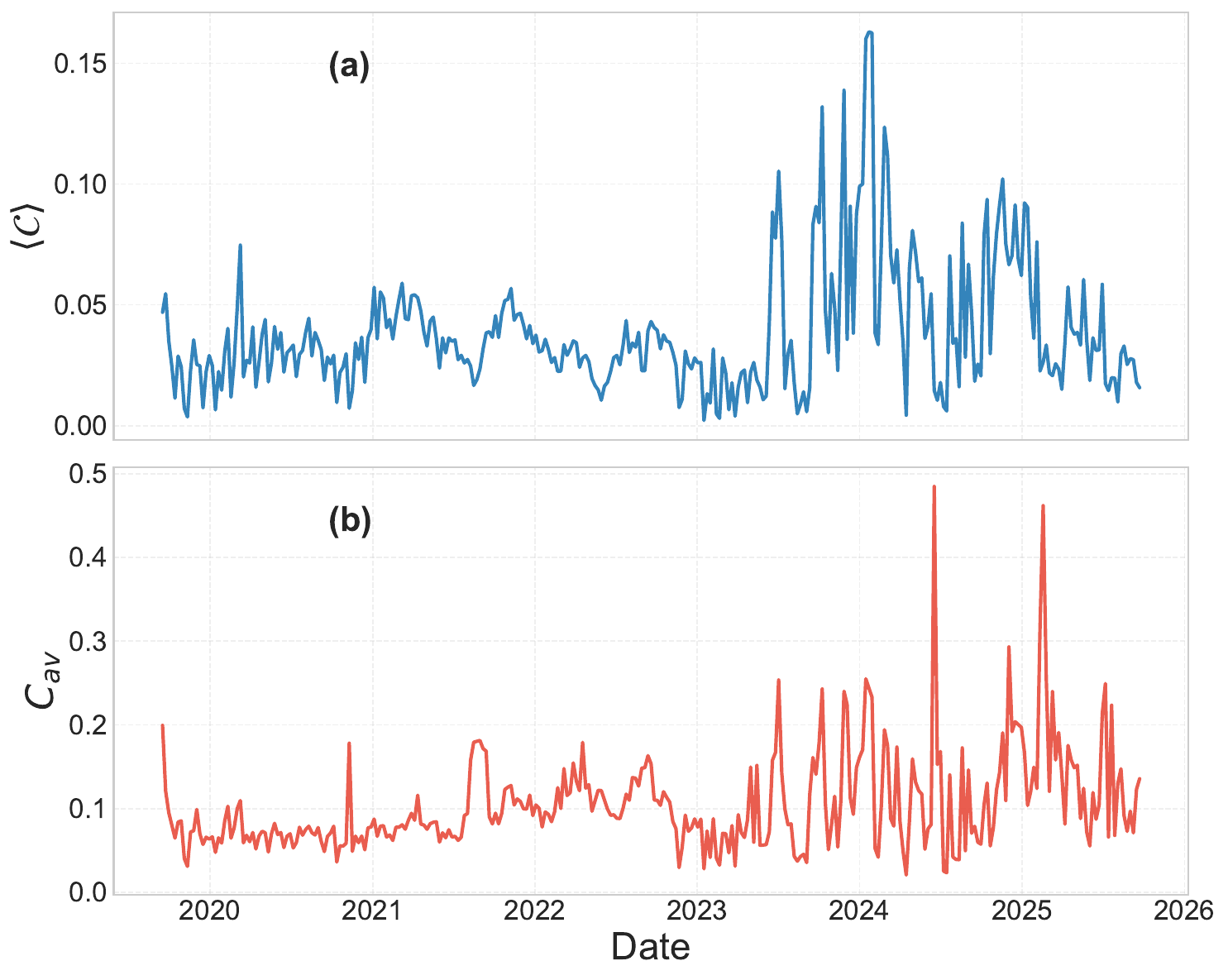}
    \caption{The evolution of the (a) average closeness centrality, $\langle \mathcal{C} \rangle$, and, (b) average clustering coefficient ,$C_{av}$, within the directed Hedera transaction network. The distinct temporal trajectories demonstrate the decoupling of global reachability from local transitivity}
    \label{fig:centrality_vs_clustering}
\end{figure}

The comparative temporal evolution of average closeness centrality, $\langle \mathcal{C} \rangle$, and the average clustering coefficient, $C_{av}$, is depicted in Figure~\ref{fig:centrality_vs_clustering}. It is worth observing that $\langle \mathcal{C} \rangle$ undergoes sharp and periodic declines during specific weeks. These localized drops reflect moments of structural stress, during which the ledger global routing capacity is constrained. In contrast, $C_{av}$ captures the density of microscopic, closed triadic trading loops, representing localized phenomena such as arbitrage cycles within a specific decentralized exchange (DEX) or isolated enterprise treasury routing. 
During periods of macroscopic network fragmentation, the connected component frequently polarizes into dense local clusters separated by severe topological bottlenecks. In these anomalous states, local clustering can remain deceptively high within these isolated pockets, even while global reachability across the network completely deteriorates. By illustrating that high local transitivity does not guarantee systemic liquidity flow, Figure~\ref{fig:centrality_vs_clustering} also justifies our decision to exclude $C_{av}$ from the composite Inefficiency Score ($\mathcal{I}$), instead relying on $\langle \mathcal{C} \rangle$ to capture the true collapse of global ledger pathways accurately.

\begin{figure}[h]
    \centering
    \includegraphics[width=\columnwidth]{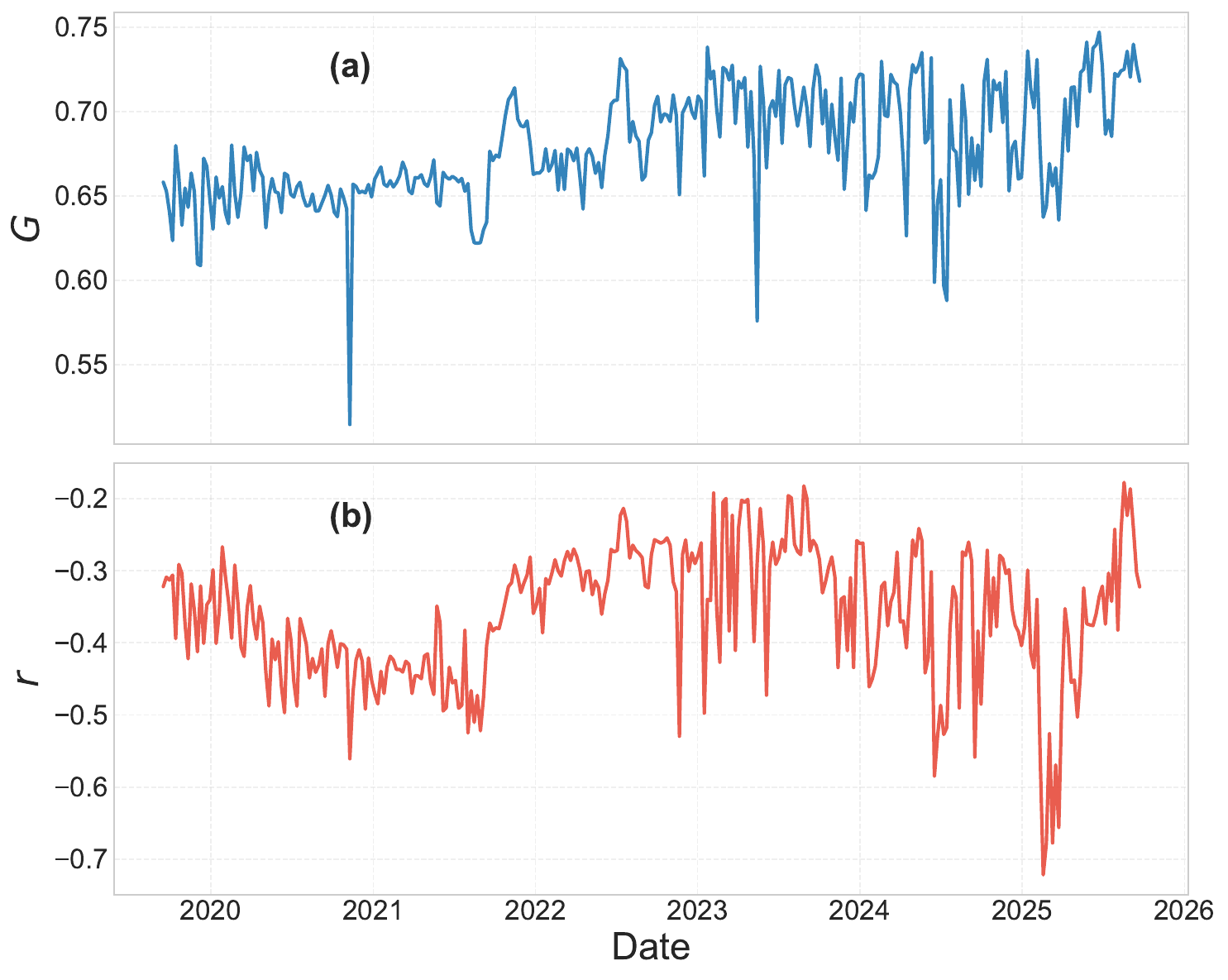}
    \caption{The time evolution of the (a) degree Gini coefficient, $G$, and, (b) degree assortativity, $r$, within the undirected version of Hedera transaction network. The persistently elevated inequality ($G$) coupled with deeply negative assortativity ($r$) characterizes a heavily centralized, disassortative hierarchy dominated by critical institutional nodes.}
    \label{fig:gini_assortativity}
\end{figure}
The temporal trajectories of the Gini coefficient, $G$, and degree assortativity, $r$, explicitly map the evolving socioeconomic hierarchy and institutional consolidation of the Hedera ledger, reported in Figure~\ref{fig:gini_assortativity}. Throughout the considered period, $G$ remains consistently high, which simply means the network connections are extremely unequal. A very small number of accounts control nearly all of the transaction traffic. This extreme centralization is topologically reinforced by the persistently negative assortativity. Given $r < 0$, the network is disassortative, so major hubs may not form a highly dense, interconnected core with one another. Apart from the connections between each other, they also act independently to serve a massive periphery of low-degree retail accounts and automated agents, producing a strict core-periphery architecture. Within the context of our anomaly detection framework, tracking $G$ and $r$ provides a diagnostic context. While our $1$D Inefficiency score ($\mathcal{I}$) isolates the resulting friction in global routing, abrupt fluctuations in $G$ and $r$ may reveal the structural catalysts of that stress. The unsupervised Machine Learning algorithm sometimes flags an anomaly exactly when these specific metrics sharply deviate. This alignment provides compelling evidence that the topological shock involves the emergence, collapse, or sudden isolation of the ledger central institutional pillars.
\begin{figure}[h]
    \centering
    \includegraphics[width=\columnwidth]{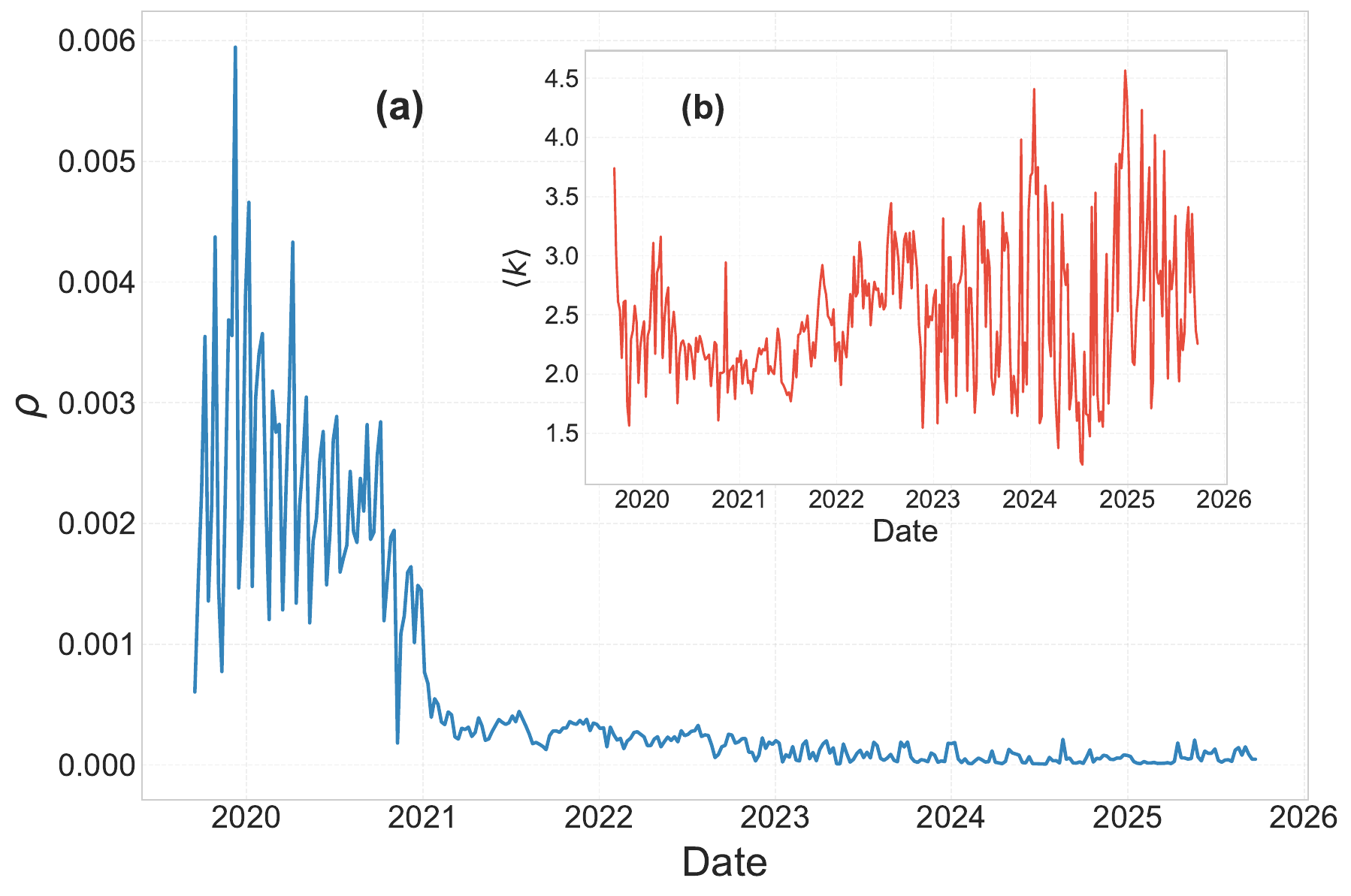}
    \caption{The evolution of (a) network density, $\rho$, and, (b) average degree, $\langle k \rangle$, associated with the directed Hedera transaction network. The steep drop in density, alongside a stable average degree, shows that the network is simply expanding as more users join. This confirms that these metrics track natural, healthy growth rather than structural stress.}
    \label{fig:density_degree}
\end{figure}
In Figure~\ref{fig:density_degree}, the network density, $\rho$, and average degree, $\langle k \rangle$, highlight the natural growth and scaling dynamics of the Hedera ledger, separating the concept of network size from structural stress. In the early stages of the network, density exhibits significant volatility. Yet, as user adoption rapidly accelerates post-2020, $\rho$ experiences a massive drop, essentially flatlining near zero. This is a standard feature of growing networks. While millions of new users join, they don't connect with everyone; they usually just interact with a small handful of popular apps and exchanges. This localized behavior is confirmed in Figure~\ref{fig:density_degree}(b), where $\langle k \rangle$ remains relatively bounded between $1.5$ and $4.5$ across the timeline. Shortly, the average user maintains a consistent transactional activity even as the overall network expands massively around them. Within the context of our anomaly detection framework, this baseline observation is relevant, as it visually establishes that density and average degree merely track network volume and user adoption, completely distinct from routing efficiency. 
This scenario motivates our subsequent Principal Component Analysis (PCA). By explicitly decoupling our $1$D Inefficiency metric from these growth-based metrics, we ensure that measurements of structural stress are not artificially distorted by the ledger natural expansion.
\begin{figure}
    \centering
    \includegraphics[width=\columnwidth]{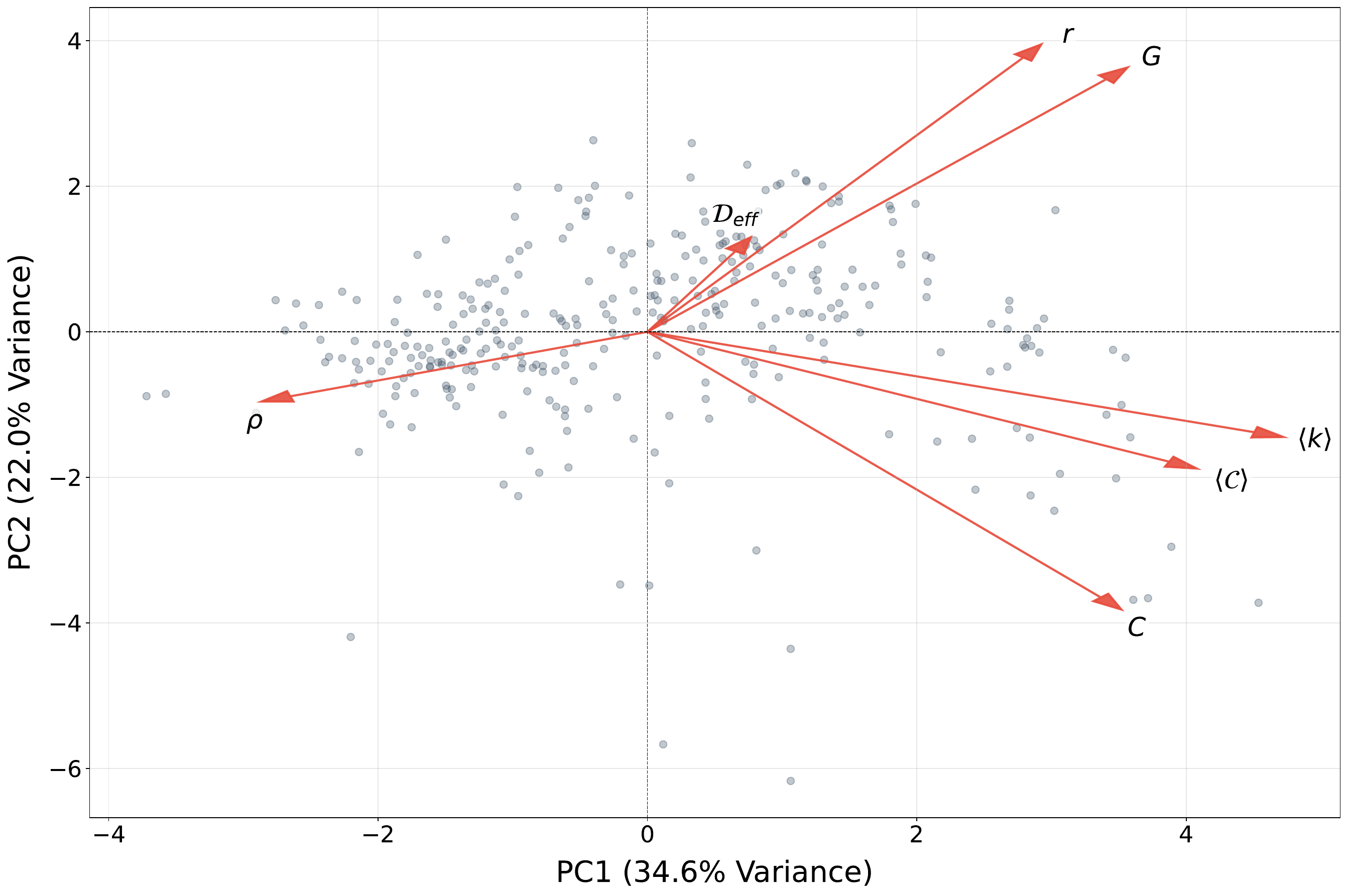}
    \caption{Principal Component Analysis (PCA) bi-plot of the Hedera network's topological phase space. The weekly structural states (grey points) are projected onto the first two principal components, capturing $56.6\%$ of the total structural variance. The feature loading vectors (red arrows) map the directional influence of the original topological metrics. Herein, the considered features are: average degree ($\langle k \rangle$), effective diameter with $90^{th}$ percentile ($\mathcal{D}_{eff}$), assortativity ($r$), Gini index ($G$), closeness centrality ($\langle \mathcal{C} \rangle$), average clustering coefficient ($C_{av}$) and density ($\rho$).}
    \label{fig:pca}
\end{figure}
DLT networks are fundamentally multidimensional complex systems whose evolution is governed by overlapping topological dynamics. To objectively identify the independent macro-states underlying the HWTN and evaluate structural redundancy among its metrics, we analyze the complete standardized topological feature space using Principal Component Analysis (PCA). Projecting the weekly network states onto the first two principal components (see Figure~\ref{fig:pca}) reveals the dominant modes of temporal structural variance, with PC1 and PC2 jointly explaining $56.6\%$ of the total variance ($34.6\%$ and $22.0\%$, respectively). The geometric arrangement of the feature loading vectors (red arrows) provides a direct representation of the latent structural constraints of the underlying network. Within this primary two-dimensional projection, the feature loading vectors organize into three apparent geometric clusters. The features corresponding to $\langle \mathcal{C} \rangle$, $\langle k \rangle$, and $C_{av}$ exhibit strong positive collinearity, suggesting coupled topological dynamics along the primary axis. Conversely, $G$, $r$, and $\mathcal{D}_{eff}$ project into a heavily divergent quadrant, representing a distinct mode of projected variance. Finally, the vector for network density ($\rho$) projects in near-perfect opposition to the primary feature cluster. However, because this $2$D plane captures only a partial variance ($56.6\%$) of the network total structural value, its visual inspection does not guarantee correlation. This dimensional limitation necessitates a formal pairwise correlation analysis to verify the independence of these structural metrics.

\begin{figure}[t]
    \centering
    \includegraphics[width=\columnwidth]{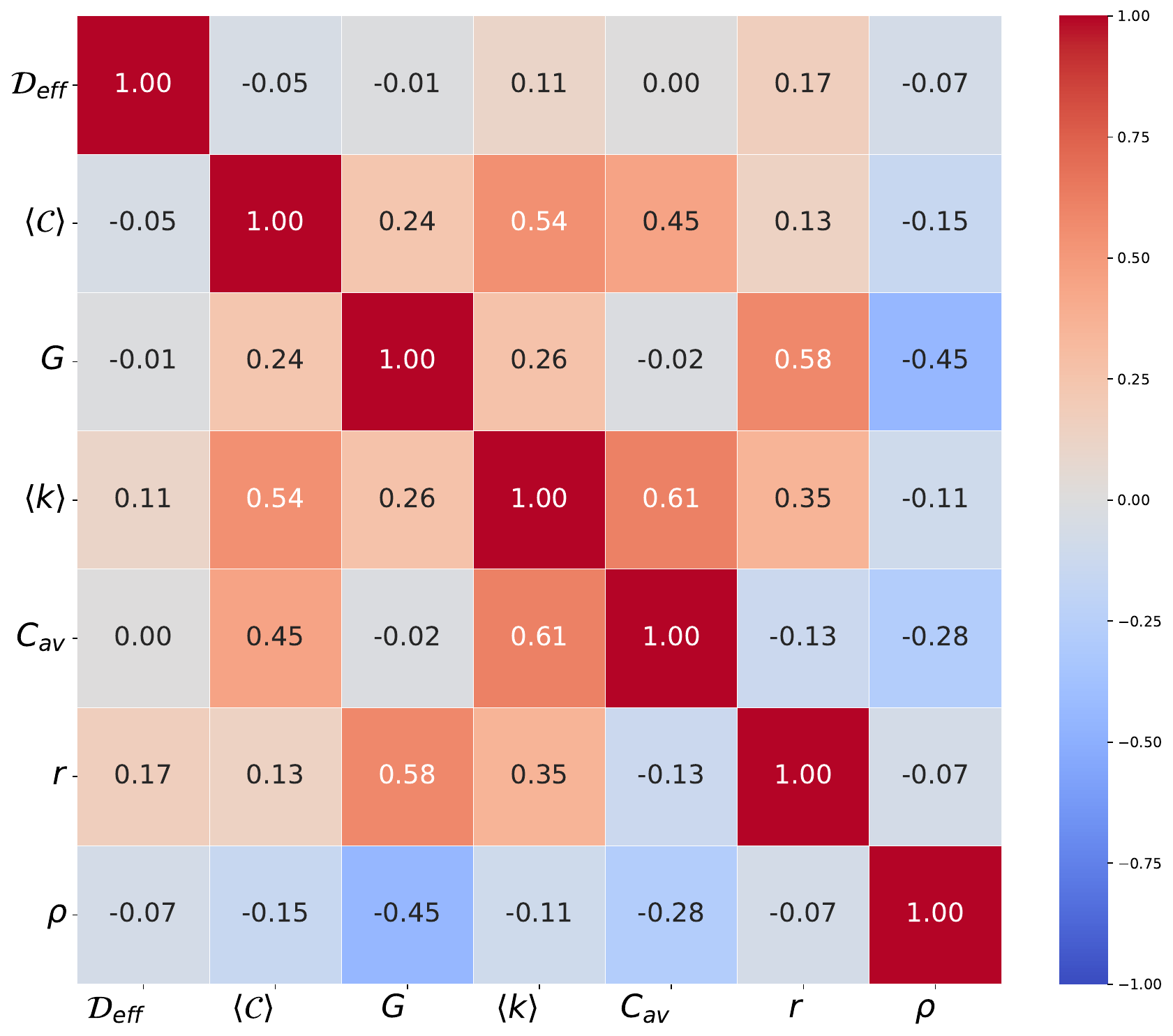}
    \caption{Pearson correlation matrix of the multiple topological measures associated with the Hedera network. The heatmap quantifies the linear covariance between macro-state variables, revealing distinct clusters of structural redundancy and orthogonality.}
    \label{fig:FC_heatmap}
\end{figure}

Then, to empirically evaluate the structural redundancy of the Hedera network, we compute a Pearson correlation matrix across the chosen topological characteristics---see Figure~\ref{fig:FC_heatmap}. The resulting heat map allows us to separate true relationships from visual artifacts present in the $2$D PCA projection. First, we examine the fundamental routing metrics, observing a moderate positive correlation between average degree ($\langle k \rangle$) and closeness centrality ($\langle \mathcal{C} \rangle$) ($r = 0.54$). In physical terms, as accounts establish more connections, on average the global pathways across the ledger become shorter, improving overall routing efficiency. Similarly, $\langle k \rangle$ correlates positively with average clustering ($C_{av}$) ($r = 0.61$), reflecting that higher overall connectivity naturally fosters dense and localized trading loops, such as user activity concentrated within a specific decentralized exchange. Furthermore, the matrix confirms that network density ($\rho$) truly possesses an inverse relationship with the broader structural features, validating its opposing direction in the PCA plot. 
Most notably, $\rho$ demonstrates a negative correlation with the Gini coefficient ($r = -0.45$). This entails that as the whole ledger grows in size and becomes sparser, the inequality of connections becomes more extreme. Also, the correlation matrix resolves a major visual limitation of the PCA. While $\mathcal{D}_{eff}$, $G$, and $r$ appear to point in the same direction in the $2$D phase space, they are not highly correlated. $\mathcal{D}_{eff}$ shows almost zero correlation with $G$ ($r = -0.01$) and only a very weak correlation with $r$ ($r = 0.17$). This is a relevant finding, as it suggests that the network extreme spatial stretch ($\mathcal{D}_{eff}$) operates independently from the network centralization ($G$). Since $\mathcal{D}_{eff}$ captures a completely distinct structural property that is not redundant with volume, density, or hierarchy metrics, it serves as an ideal component for our Inefficiency Metric.

Both the PCA feature space and the Pearson correlation matrix provide strict empirical justification for the variables chosen for our proposed Inefficiency Score ($\mathcal{I}$). While the data reveals several independent structural dimensions within the network, it is crucial to distinguish between a network structural organization and its actual routing capacity. For example, effective diameter ($\mathcal{D}_{eff}$) and the Gini coefficient ($G$) are independent ($r = -0.01$). However, we reject this pairing because $G$ strictly measures connection inequality, which does not directly indicate systemic routing failure. Similarly, while assortativity ($r$) and closeness centrality ($\langle \mathcal{C} \rangle$) are highly independent ($r = 0.13$), they measure fundamentally different physical properties. Assortativity describes how nodes group together based on their similarity, which in this case is related to their degree, but it cannot describe the macroscopic spatial boundaries of the ledger. To construct a meaningful and deterministic measure of network stress, we must isolate the specific physical tension between the ledger global expansion and its internal routing core. Incorporating additional variables like $G$ or $r$ would dilute the physical interpretability of the metric, as they capture hierarchy and mixing patterns rather than routing degradation. By restricting the metric strictly to $\mathcal{D}_{eff}$ and $\langle \mathcal{C} \rangle$, we ensure the Inefficiency Score remains a highly focused, minimally sufficient descriptor of the fundamental trade-off between the network extreme spatial stretch and its global internal reachability.

\begin{figure}[t]
    \centering
    \includegraphics[width=\columnwidth]{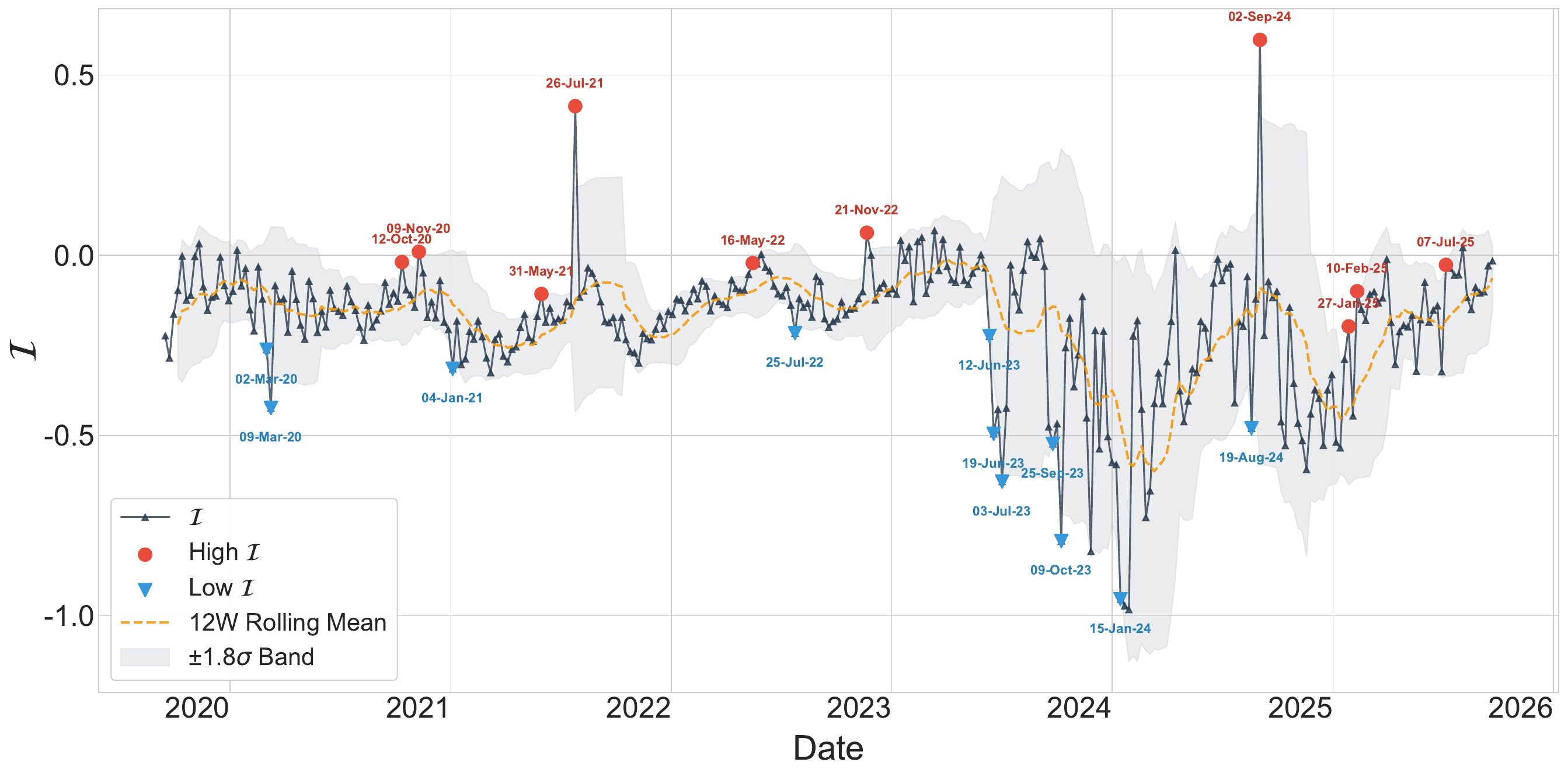}
    \caption{Temporal evolution of the Hedera network's topological inefficiency score ($\mathcal{I}$) with rolling Z-score anomaly detection. A 12-week rolling baseline and adaptive $\pm 1.8\sigma$ threshold identify localized structural deviations from normal network behavior.}
    \label{fig:IS_rolling}
\end{figure}

To systematically isolate periods of anomalous structural behavior, we evaluated the Inefficiency Metric ($\mathcal{I}$) against a 12-week rolling mean ($\mu$). As detailed in Table~\ref{tab:IS_events}, the first category of topological anomalies, characterized by high inefficiency ($\mathcal{I} > \mu + 1.8\sigma$), reflects periods of significant network stretching and decentralized complexity. These events occur when the effective diameter expands while closeness centrality decreases, indicating that transaction pathways have become longer and less centralized. Historically, these events align with times when the Hedera network undertook more complex operations. For example, the launch of the Hedera Token Service (HTS) or the minting of early NFTs naturally forced the network to utilize longer, multi-step smart contract pathways. Interestingly, macroeconomic shocks that destroy centralized exchanges, such as the collapses of Terra/LUNA and FTX, also trigger these high metric values. When central hubs fail, users are forced to route their funds through complex decentralized bridges and self-custody wallets, effectively stretching the network topology.
In contrast, low inefficiency states ($\mathcal{I} < \mu - 1.8\sigma$) indicate periods when the network shrinks and becomes highly compact. The network structure simplifies dramatically, with the vast majority of users connecting directly to a few massive central hubs. This topological shrinking usually happens during a `flight to centralization'. For example, during extreme market panics like the COVID-19 `Black Thursday' crash, users abandon complex decentralized applications and instead route their funds directly to large exchanges to liquidate. We also observe this network compaction during major institutional convergence. When Spot Bitcoin ETFs were approved, or when native HBAR staking was introduced, millions of user accounts linked directly to a small number of massive custodial hubs or network nodes.

Ultimately, the dichotomy between high and low inefficiency demonstrates that the $\mathcal{I}$ acts as a structural indicator for capital routing, rather than a simple directional price indicator. High values reflect widespread, peer-to-peer decentralization, whereas low values reflect network compaction around massive central hubs. Tracking these anomalies chronologically illustrates how the Hedera network adapts to different market conditions. Following the decentralized complexity forced by the 2022 centralized exchange collapses, the network anomalies throughout 2023 and 2024 were overwhelmingly characterized by structural compaction (low $\mathcal{I}$). Driven by the introduction of native staking and institutional convergence (such as the approval of Spot Bitcoin ETFs), capital consolidated around a few highly regulated custodial hubs, triggering repeated low-inefficiency states. However, the data in 2025 shows a return to extreme high-inefficiency anomalies. This realistic structural expansion aligns with the deployment of advanced, multi-hop enterprise infrastructure, confirming that the network stretched once again to accommodate sovereign CBDC pilots and complex real-world asset (RWA) tokenization.

\begin{table*}
\centering
\caption{Detected high- and low-inefficiency windows in the Hedera network.}
\label{tab:IS_events}
\scriptsize

\begin{tabular}{ll}
\toprule
\textbf{Window} & \textbf{Associated Events} \\
\midrule

\multicolumn{2}{c}{\textbf{High Inefficiency}} \\
\midrule

Oct 12--18, 2020 & PayPal crypto integration and preparation for Hedera Token Service (HTS) \\ 
Nov 9--15, 2020 & COVID-19 vaccine by Pfizer/BioNTech and HTS Previewnet \\ 
May 31--Jun 6, 2021 & First Hedera NFTs and ESG backlash against PoW networks and market recovery after cryptocurrency crash \\
Jul 26--Aug 1, 2021 & China Evergrande Crisis and Market consolidation and heavy correlation with tech equities \\ 
May 16--22, 2022 & Terra/LUNA collapse and DeFi systemic contagion \\
Nov 21--27, 2022 & collapse of the FTX and Joget and HBAR Foundation partnership \\
Sep 2--8, 2024 & Macro hesitation and pre-NFP market consolidation, starting phase of Project-Hiero \\ 
Jan 27--Feb 2, 2025 & BTC breaks 100k, and Hedera focus on AI/DeFi \\
Feb 10--16, 2025 & phase before HederaCon 2025 annoucement, ongoing Hiero Codebase Transition \\ 
Jul 7--13, 2025 & ``Crypto Week", RBA CBDC pilot, release of Hedera CLI and Lloyds Bank RWA trade \\ 
\midrule

\multicolumn{2}{c}{\textbf{Low Inefficiency}} \\
\midrule

Mar 2--8, 2020 & COVID-19 panic and launch of Hedera Consensus Service (HCS) \\ 
Mar 9--15, 2020 & ``Black Thursday" crash and COVID-19 pandemic \\
Jan 4--10, 2021 & Institutional FOMO, Tesla BTC purchase, BTC price surge, high activity in Hedera and early NFT craze \\
Jul 25--31, 2022 & Introduction of Native staking and delegation of HBAR, Hedera open-source pivot \\ 
Jun 12--18, 2023 & SEC lawsuits against major centralized exchanges, active account and TVL growth of Hedera \\ 
Jun 19--25, 2023 & BlackRock ETF filing and TradFi market convergence \\
Jul 3--9, 2023 & Ripple partial legal victory, altcoin relief rally and XRP Price Surge \\
Sep 25--Oct 1, 2023 & Post activity of Stablecoin Studio Launch, ServiceNow ESG initiative, high and stable USDT transfers on the TRON blockchain \\ 
Oct 9--15, 2023 & active ServiceNow integration, WalletConnect integration initiative for security and ``Hello Future" event in LA \\ 
Jan 15--21, 2024 & Spot BTC ETFs launch, Hitachi joins Council and EQTY AI rollout \\
Aug 19--25, 2024 & HelloFuture Hackathon Conclusion and weak recovery from ``Black Monday" on August 5 \\

\bottomrule
\end{tabular}
\end{table*}

\begin{figure}[t]
    \centering
    \includegraphics[width=\columnwidth]{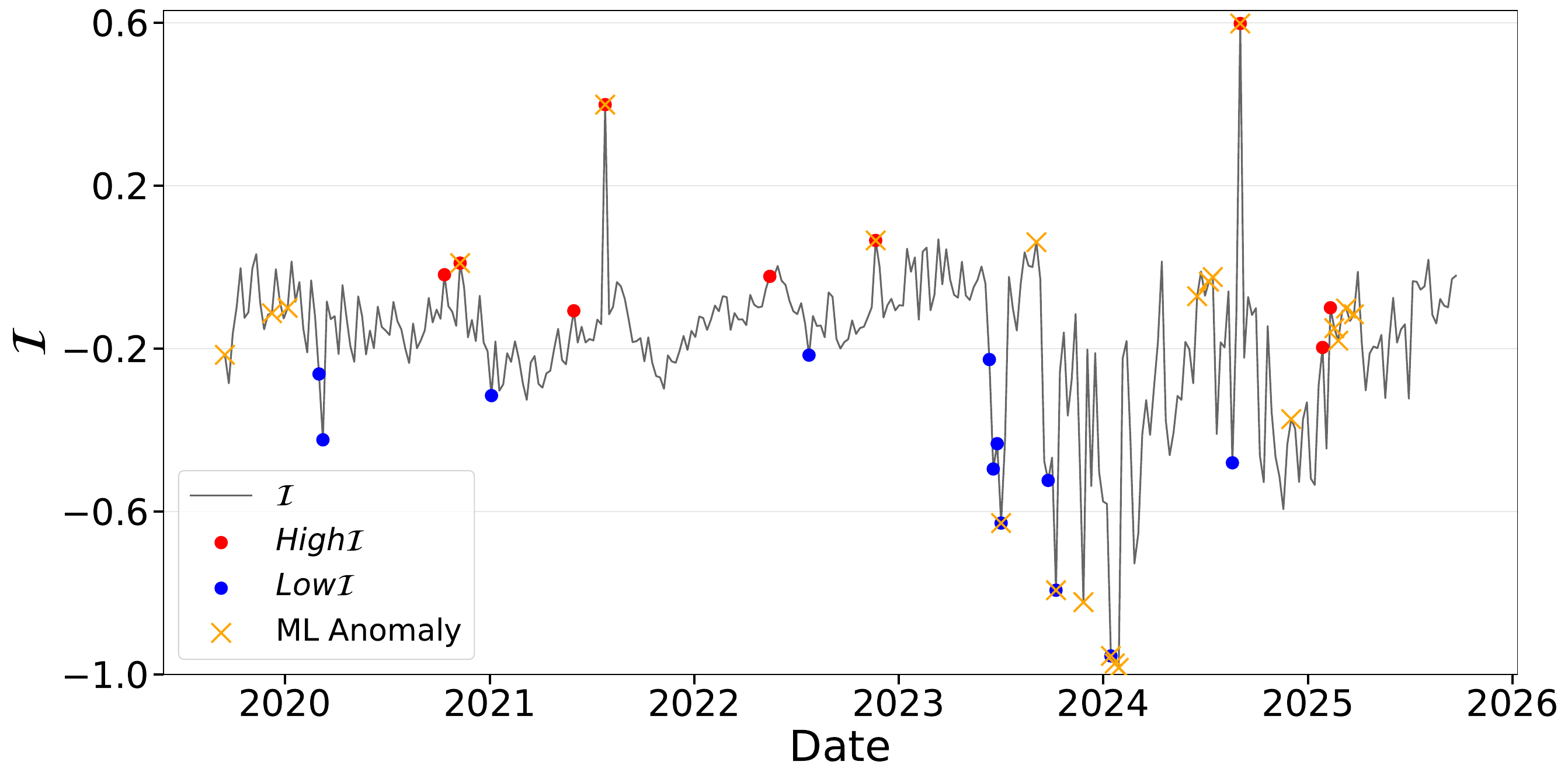}
    \caption{Unsupervised machine learning detection of acute topological anomalies using an Isolation Forest algorithm. Orange crosses denote severe structural outliers autonomously identified across a seven-dimensional feature space. To contextualize these transient events, the ML-derived anomalies are superimposed over the Inefficiency Score ($\mathcal{I}$), highlighted in red and blue circles, as shown in Fig. \ref{fig:IS_rolling}.} 
    \label{fig:ML_rolling}
\end{figure}

%

\begin{table*}
\centering
\caption{Isolation Forest anomaly windows and their associated real-world events. The weeks highlighted in bold have been also identified using Inefficiency Metric.}
\label{tab:ML_events}
\scriptsize

\begin{tabular}{ll}
\toprule
\textbf{Window} & \textbf{Real-World Events} \\
\midrule
Sep 16--22, 2019 & Initial phase \\ %
Dec 9--15, 2019 & No significant events \\ %
Jan 6--12, 2020 & Ethereum DeFi TVL hits \$1B \\ 
\textbf{Nov 9--15, 2020} & COVID-19 vaccine by Pfizer/BioNTech and HTS Previewnet \\ 
\textbf{Jul 26--Aug 1, 2021} & China Evergrande Crisis and Market consolidation and heavy correlation with tech equities \\ 
\textbf{Nov 21--27, 2022} & collapse of the FTX and Joget and HBAR Foundation partnership \\
\textbf{Jul 3--9, 2023} & Ripple partial legal victory, altcoin relief rally and XRP Price Surge \\
Sep 4--10, 2023 & BTC price drop \\
\textbf{Oct 9--15, 2023} & active ServiceNow integration, WalletConnect integration initiative for security and ``Hello Future" event in LA \\ 
Nov 27--Dec 3, 2023 & Bullish surge of BTC \\
\textbf{Jan 15--21, 2024} & Spot BTC ETFs launch, Hitachi joins Council and EQTY AI rollout \\
Jan 22--28, 2024 & Heavy Grayscale outflows vs. BlackRock inflows, Massive Transaction and performance growth in Hedera \\
Jan 29--Feb 4, 2024 & ETF liquidity rebalances \\ 
Jun 17--23, 2024 & SEC drops ETH 2.0 probe \\ 
Jul 8--14, 2024 & No significant events \\
Jul 15--21, 2024 & Hello Future Hackathon 2024 from July 22 \\
\textbf{Sep 2--8, 2024} & Macro hesitation and pre-NFP market consolidation, starting phase of Project-Hiero \\ 
Dec 2--8, 2024 & No significant events \\
Feb 17--23, 2025 & No significant events \\ 
Feb 24--Mar 2, 2025 & HederaCon 2025 BTC and largest crypto exchange theft in Bybit \\ 
Mar 10--16, 2025 & No significant events \\
Mar 24--30, 2025 & High Network Activity \\
\bottomrule
\end{tabular}
\end{table*}

Cross-referencing the unsupervised machine learning anomalies with the $\mathcal{I}$ provides a critical validation mechanism for our methodology. As detailed in Table~\ref{tab:ML_events}, the Isolation Forest identifies $22$ multi-dimensional anomalies across the timeline. Most importantly, $7$ of these dates align with the statistical extremes of $\mathcal{I}$ (highlighted in bold). These overlapping periods correspond directly to universally recognized macroeconomic shocks and major ecosystem events, such as the collapse of FTX (November 2022) and the launch of Spot BTC ETFs (January 2024). When both the high-dimensional machine learning model and our deterministic $\mathcal{I}$ metric trigger simultaneously, it indicates that the network is experiencing profound structural stress, warping across all measured dimensions. Ultimately, the Isolation Forest serves as robust proof that the peaks of the $\mathcal{I}$ accurately target the most severe disruptions in the network's history.
However, an analysis of the $15$ dates flagged exclusively by the Isolation Forest highlights the practical limitations of using purely high-dimensional anomaly detection for this specific analysis. The machine learning model frequently flags periods associated with `No significant events', localized hackathons (e.g., July 2024), or routine liquidity rebalances (e.g., January 2024). From a network science perspective, this behavior suggests that the Isolation Forest is overly sensitive to endogenous micro-noise. The algorithm evaluates seven topological dimensions simultaneously, making it highly sensitive to isolated fluctuations. Consequently, a strictly localized event—such as a single institution reorganizing its wallets—can trigger the anomaly threshold. This occurs even when the network global capital routing pathways remain completely undisturbed.
This divergence demonstrates why the proposed Inefficiency Metric ($\mathcal{I}$) can be a reliable and practical tool for identifying genuine structural changes. High-dimensional machine learning models often operate as black boxes lacking physical interpretability and directionality; they can identify a localized hackathon and a global market collapse as identical outliers. By intentionally restricting the $\mathcal{I}$ metric strictly to effective diameter ($\mathcal{D}_{eff}$) and closeness centrality ($\langle \mathcal{C} \rangle$), we apply physical parsimony to filter out the transient dimensional noise of the ledger. This ensures that the metric only flags an anomaly when the macroscopic routing behavior of the users and institutions on the network undergoes a severe, directional shift.

\section{Conclusion}\label{sec:conclusion}
In this work, we presented a topological analysis of the Hedera transaction network, moving beyond standard volume-based metrics to isolate the true structural drivers of network stress. 
By utilizing Principal Component Analysis (PCA) and Pearson correlation, we demonstrated that metrics such as network density ($\rho$) and average degree ($\langle k \rangle$) primarily capture natural network scaling, while connection inequality ($G$) and assortativity ($r$) describe hierarchical organization rather than routing capacity. 
Thus, we proposed the Inefficiency Metric ($\mathcal{I}$), i.e. a deterministic structural indicator built on effective diameter ($\mathcal{D}_{eff}$) and closeness centrality ($\langle \mathcal{C} \rangle$). This parsimonious formulation successfully captures the physical tension between macroscopic network expansion and internal routing accessibility, without being artificially distorted by the ledger natural growth.
The application of the metric across six years of transaction data reveals that the Hedera network experiences distinct structural shifts driven by broader market forces. High-inefficiency anomalies act as an indicator of decentralized complexity, often occurring during periods of systemic expansion (such as the integration of multi-hop smart contracts) or when failures of centralized intermediaries (such as the Terra/LUNA and FTX collapses) force users into sprawling, self-custodial routing pathways. Conversely, low-inefficiency anomalies highlight periods of network compaction. These highly centralized topological states consistently emerge when transaction activity collapses into core network hubs amid market stress, or during phases of institutional consolidation, such as the rollout of native staking and the introduction of Spot Bitcoin ETFs.
To validate these findings, we cross-referenced our deterministic $1$D metric against a $7$-dimensional, unsupervised Isolation Forest algorithm. While the machine learning model identified $22$ multi-dimensional anomalies, it proved highly sensitive to localized, structurally insignificant micro-events. However, the $7$ periods where the two models aligned demonstrate that extreme spikes in $\mathcal{I}$ accurately capture periods of severe structural stress. This divergence highlights a distinct advantage of the proposed methodology, i.e. the focused $\mathcal{I}$ metric successfully filters out transient dimensional noise to isolate significant and network-wide structural changes.
Ultimately, this work suggests the Inefficiency Metric ($\mathcal{I}$) can be effectively adopted as structural indicator for Distributed Ledger Technologies. By decoupling topological stress from natural network scaling, this framework provides a clear physical measure. Also, it offers a data-driven lens to evaluate how real-world market dynamics, extreme volatility, and institutional adoption fundamentally alter the routing architecture of decentralized economies.
At the same time, we consider testing its performance on other networks essential to corroborating its relevance, and therefore leave this as a fundamental direction for future development.
\bibliography{references}

\end{document}